\documentstyle[prd,aps,psfig]{revtex}
\begin{document}
\draft
\title{Cosmological implications of old galaxies at high redshifts}
\author{J. A. S. Lima\thanks{Electronic address: limajas@dfte.ufrn.br} and 
J. S. Alcaniz\thanks{Electronic address: alcaniz@dfte.ufrn.br}}
\address{Departamento de F\'{\i}sica, Universidade Federal do Rio
Grande do Norte, C.P. 1641 59072-970 Natal RN Brazil}
\maketitle
\begin{abstract}
Old high-$z$ galaxies are important tools for understanding the structure
formation problem  and may become the key to determine the ultimate fate of
the Universe. In this {\it letter}, the inferred ages of the three oldest
galaxies at high redshifts reported in the literature are used to constrain 
the first epoch of galaxy formation and to reanalyse the high-z time scale
crisis. The lower limits on the formation redshift $z_f$ depends on the
quantity of  cold dark matter in the Universe. In particular, if $\Omega_m
\geq 0.37$ these galaxies are not formed in FRW cosmologies with no dark
energy. This result is in line with the Supernovae type Ia measurements which
suggest that the bulk of energy in the Universe is repulsive and appears like
an unknown form of dark energy component. In a complementar analysis, unlike
recent claims favoring the end of the age problem, it is shown that the
Einstein-de Sitter model is excluded at high-z by $\sim 3\sigma$. 
\end{abstract}

\vspace{0.5cm}

The existence of old high-redshift galaxies (OHRGs) provides one of  the best
methods for constraining the age of the Universe \cite{krauss,alcaniz}, and,
in a similar vein, may be an important key for determining the first epoch of
galaxy formation. Recently, different groups [3-7] announced the discoveries
of three extremely red radio galaxies at $z=1.175$ (3C 65), $z=1.55$ (53W091)
and $z=1.43$ (53W069) with a minimal  stellar age of 4.0 Gyr, 3.0 Gyr and 4.0
Gyr, respectively. These discoveries accentuated even further the already
classical ``age crisis'' and  gave  rise to a new variant of this problem,
which could be named the high-$z$ time scale crisis: the underestimated ages
of these galaxies  contradict the predictions of the standard Einstein-de
Sitter model for values of  $h > 0.45$ \cite{krauss}. For comparison, if one
considers  $h = H_o/100{\rm{Km s^{-1}Mpc^{-1}}} =  0.6$, the age of the
Universe predicted by the Einstein-de Sitter CDM model at redshifts $z=1.175$
and $z=1.55$ is $t_{z}=3.35$ Gyr and $t_{z} = 2.5$ Gyr, respectively. Such
discrepancies become even larger if one takes into account the radio galaxy at
$z=1.43$ with a minimal stellar age of 4.0 Gyr. In this case, the predicted
age is $t_{z}=2.85$ Gyr, which yields an age difference greather than 1 Gyr.
It is still worth noticing that such estimates do not include an incubation
time of at least 0.5 - 1 Gyr, which would be more in accordance to the
chemo-dynamical evolution models for such objects\cite{jime}. As it appears,
the problem raised by the ages of these galaxies to the standard CDM model is
a strong indication that OHRGs may also constrain appreciably the formation
redshift ($z_f$), i.e., the epoch where the first structures were actually
formed. In brief, this is the aim of this letter. In what follows, robust
lower limits on $z_f$ are derived for  open cold dark matter (OCDM) and flat
$\Lambda$CDM models from the estimated ages of the quoted  old high-$z$
galaxies. For completeness, we also present a new quantitative approach for
analysing the high-$z$ time scale crisis.

Many attempts for determining the first epoch of galaxy formation have been
done in  the past, and the overall conclusion is that  the major period of
galaxy formation lies in the redshift interval $1 \leq z \leq 3$, that is, at
relatively low redshifts \cite{cowie,zep}. Such results are consistent with
the simplest cold dark matter (CDM) scenario for structure formation, although
its modified version ($\Lambda$CDM) gave a more natural explanation for the
excess of power observed in the galaxy distribution \cite{maddox,efsta}.  In
spite of that, recent findings of galaxy candidates at $z > 4.0$ [13-15]
suggest that our Universe contains more collapsed objects than believed some
few years ago, and all this activity at high redshifts cannot be easily
accommodated in the CDM scenarios \cite{kas}. Since such studies do not
exclude the possibility that rare events of galaxy formation took place at
higher redshifts, the very beginning of the structure formation process
remains as one of the most challenging problems in modern cosmology.

Let us first recall that for the class of passively evolving elliptic radio
galaxies, almost all  amount  of gas is  believed to be processed into stars
in a single episode of star formation, in such a way that the assumption of an
 instantaneous burst is considered a good approximation for modeling their
evolution. For lookback time calculations, an instantaneous burst of star 
formation means that the age of these OHRGs can be expressed as being almost
exactly the time taken by the Universe to evolve from $z_f$ to the observed
redshift $z_{obs}$. In the framework of  Friedmann-Robertson-Walker (FRW)
models with cosmological constant such condition can be translated as    
\begin{equation}   t(z_{obs}) - t(z_f) = {H_{o}}^{-1}
\int_{(1+z_f)^{-1}}^{(1+z_{obs})^{-1}} {dx \over \sqrt{1 -  \Omega_m +
\Omega_m x^{-1} + \Omega_{\Lambda}(x^2 - 1)} } \geq t_g \quad,  
\end{equation}
where $\Omega_m$ and  $\Omega_{\Lambda}$ stand for the present-day matter and
vacuum  density parameters, respectively. The inequality signal on the r.h.s.
of the above expression comes from the fact that the Universe is older than or
at least  has the same age of any observed structure. Since this natural
argument also holds for any time interval, a finite value for the redshift
$z_f$ provides the lower bound for the galaxy  formation allowed by the aged
object located at $z_{obs}$. Models for which $z_f \rightarrow \infty$ are
clearly incompatible with the existence of the specific galaxy, being ruled
out in a natural way. 

Before discussing the resulting diagrams, an important  point of principle
should be stressed. To assure the robustness of the limits on $z_f$, we addopt
(1) the  minimal value for the Hubble parameter,  and (2) the underestimated
age for all OHRGs. Both conditions are almost self-explanatory. First, as we
know, the smaller the value of $H_{o}$, the larger the age predicted by the
model and, second, objects with smaller ages are theoretically more easily
accommodated, thereby guaranteeing that the models are always favored in the
present estimates. For the Hubble parameter we consider the  value obtained by
the HST Key project which is in agreement with other independent estimates
\cite{giov}, i.e., the  round number value $H_{o} = 60\rm{km/sec/Mpc}$
\cite{freed,freed1}. Indeed, we are being rather conservative since this lower
limit was recently updated to nearly $10\%$ of accuracy ($h = 0.70 \pm 0.07$,
$1\sigma$) by Friedman \cite{freed1}, and the data from SNe also point
consistently to $h > 0.6$ or even higher \cite{perlmutter,riess}.

In Fig. 1a we show the $\Omega_m - z_f$ plane allowed by the  existence of
these OHRG's for OCDM models. The shadowed
horizontal region corresponds to the observed range $\Omega_{\rm{m}} = 0.2 -
0.4$ \cite{dekel}, which is used to fix the lower limits on $z_f$. As should
be  physically expected, if the matter contribution increases, a larger value
of $z_f$ is  required in order to account for the existence of these OHRGs
within these models. Conversely, for each object, the absolute minimal value
of $z_f$ is obtained for an empty universe ($\Omega_{\rm{m}} \rightarrow 0$).
In the observed range of $\Omega_m$ the allowed values for the formation
redshift are unexpectedly high. For example, by considering $\Omega_{\rm{m}} =
0.3$, as indicated from dynamic estimates on scales up to about 2$h^{-1}$
Mpc \cite{calb}, the ages of 3C 65, 53W091 and 53W069 provide, respectively,
$z_f \geq 6.3$, $z_f \geq 10.5$ and $z_f \geq 18$. Such values suggest that
these galaxies were formed about 12.5 Gyr ago, or by considering the most
recent lower limits for the age of the Universe \cite{carr}, that such objects
were formed when the Universe was $\sim$ 1.0 Gyr old. However, since almost
all the age of the Universe is at low redshifts ($z = 0 - 2$), these galaxies
may have been formed nearly at the same epoch, regardless of their constraints
on the redshift space. 

Figure 1b  shows similar plots for flat $\Lambda$CDM models. In this case,
the effect of the  equation of state associated to the ``vacuum medium" ($p_v
=-\rho_v$) is to accelerate the cosmic expansion. In particular, this means
that the lookback time between $z_{obs}$ and $z_f$ is larger than in the
standard scenario and, therefore, the galaxy formation process may start
relatively late in comparison to the corresponding OCDM model. For example, if
the density parameter is around the central value, $\Omega_{\rm{m}}= 0.3$, the
ages of 3C 65, 53W091 and 53W069 restrict the formation redshift  by  $z_f
\geq 3.6$, $z_f \geq 5.2$ and $z_f \geq 5.8$, respectively. The value of $z_f$
is also  proportional to the quantity of dark matter. As one may check, in
the limiting case of a universe dominated only by dark energy, the lower limit
is $z_f \geq 2$. It is worth noticing that for $\Omega_{\rm{m}} \geq
0.37$, the lower limit inferred from the age of 59W069 is $z_f \rightarrow
\infty$. This result is consistent with recent studies based on the
age-redshift relation\cite{alcaniz,alcaniz1}, and means that the standard
cosmological model  with $\Omega_{\rm{m}} \geq 0.37$ and  $h \geq 0.6$ is
(beyond doubt) incompatible with the existence of this galaxy.  We also stress
that the  present constraints on the formation redshift are indeed rather
conservative since the lower limit on $H_o$ has been considered in all the
estimates. Finally, we stress that these lower limits on $z_f$ provide a new
theoretical evidence that galaxies are not uncommum objects at very large
redshifts, say, at $z > 5$, and also reinforce the interest on the
observational search for galaxies and other collapsed objects within the
redshift interval $5 \leq z \leq 10$ (For a discussion about the influence of
an arbitrary equation of state $p_x = \omega \rho_x$ ($\omega \geq -1$) on the
redshift formation, see \cite{alcaniz2}).

Another important implication of age dating of high-$z$ objects is on the
expanding age of the Universe. As before, a crucial point is 
concerned  to the measurements the Hubble parameter. Recently, Sandage
{\it et al.} \cite{Sand} argued for the end of the age  crisis by concluding
that {\it there is no time scale crisis in cosmology   if $H_o = 55 \pm 5{\rm
kms^{-1}Mpc^{-1}}$} as inferred by the Hubble diagram of  52 fiducial SNe Ia.
For $h = 0.55 \pm 0.05$ and by taking $t_o =12 \pm 1$ (1$\sigma$) as the
median value for the estimated ages of globular clusters \cite{freed1}, we
find  $H_{o}t_{o} = 0.67 \pm 0.09$ which is clearly compatible with the 
prediction of the Einstein-de Sitter model ($t_o = \frac{2}{3}H_{o}^{-1}$).
Moreover, even by adding 1 Gyr as incubation time \cite{Pee} we find
$H_{o}t_{o} = 0.73 \pm 0.1$ which is also compatible with the standard flat
CDM model. However, what can be said about a similar analysis at high
redshift?

In Fig. 2a we assume 0.5 Gyr as incubation time for the LBDS 53W069 and show
the dimensionless product $H_ot_z$ as a function of $\Omega_{\rm{m}}$. Two
different cases  are illustrated: the standard CDM and the flat $\Lambda$CDM
models.  Dotted lines indicate the $\pm 2\sigma$ limits of the age-parameter
by considering $h = 0.55 \pm 0.05$ whereas dashed lines indicate the $\pm
2\sigma$ limits for $h = 0.7 \pm 0.1$, the value obtained by the {\it HST Key
Project} \cite{freed} and by other independent estimates 
\cite{giov,perlmutter}. At this redshift ($z = 1.43$), the prediction of the
flat  matter-dominated model is $H_o t_z \leq 0.17$ whereas for $h = 0.55 \pm
0.05$ this galaxy yields $H_o t_z = 0.22 \pm 0.03$, which rules out the
Einstein-de Sitter case by $\sim$ 2 standard deviations. In the case of the
LBDS 53W091 (at $z = 1.55$), the prediction of the Einstein-de Sitter model is
$H_o t_z \leq 0.16$, and similar analysis provides $H_o t_z = 0.20 \pm 0.03$.
As expected, such discrepancies become even larger if one considers $h = 0.7
\pm 0.1$. In this case, we find for the LBDS 53W091 and LBDS 53W069 $H_o t_z =
0.29 \pm 0.04$ and $H_o t_z = 0.25 \pm 0.05$, which excludes the Einstein-de
Sitter model by $2\sigma$ and $3\sigma$, respectively. Therefore, if the
estimated age of these objects are correct, we may conclude, at light of the
above analysis that there is a high-$z$ time scale crisis in the standard
cosmology, even by considering low values for the Hubble parameter. Naturally,
such a result is in agreement with the latest measurements from Supernovae
type Ia which point to the existence of a repulsive energy component
dominating the bulk of matter-energy in the Universe, thereby implying that
the Universe is necessarely older than the age predicted by the Einstein-de
Sitter model. 

\begin{center} 
{\bf Acknowledgements} 
\end{center}   
This research was partially supported by the Conselho Nacional de
Desenvolvimento Cient\'{\i}fico e  Tecnol\'{o}gico (CNPq), Pronex/FINEP.
(no. 41.96.0908.00) and FAPESP (no. 00/06695-0)


\newpage

\begin{figure} 
\vspace{.2in}  
\centerline{\psfig{figure=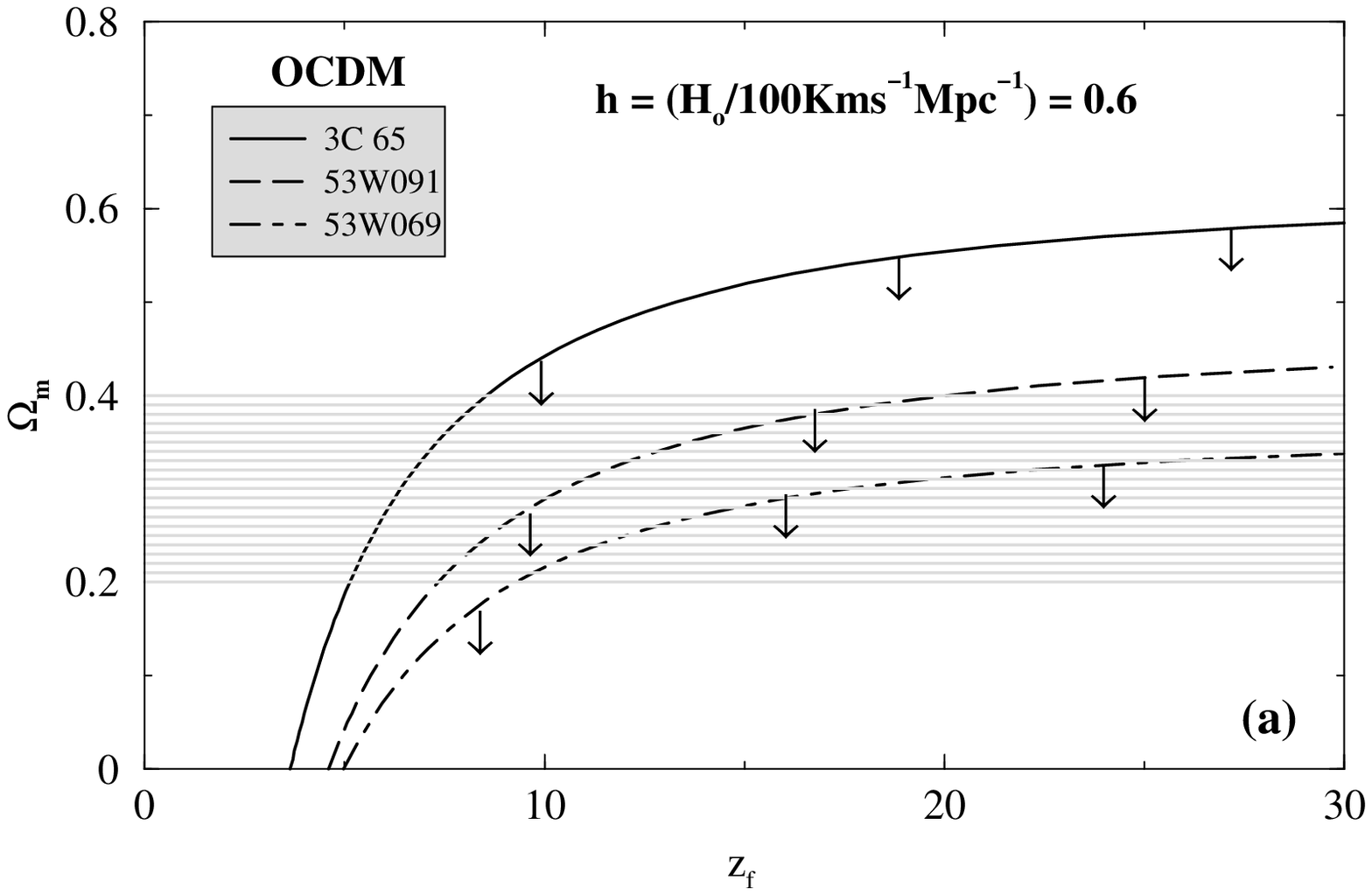,width=3.0truein,height=3.0truein}  
\psfig{figure=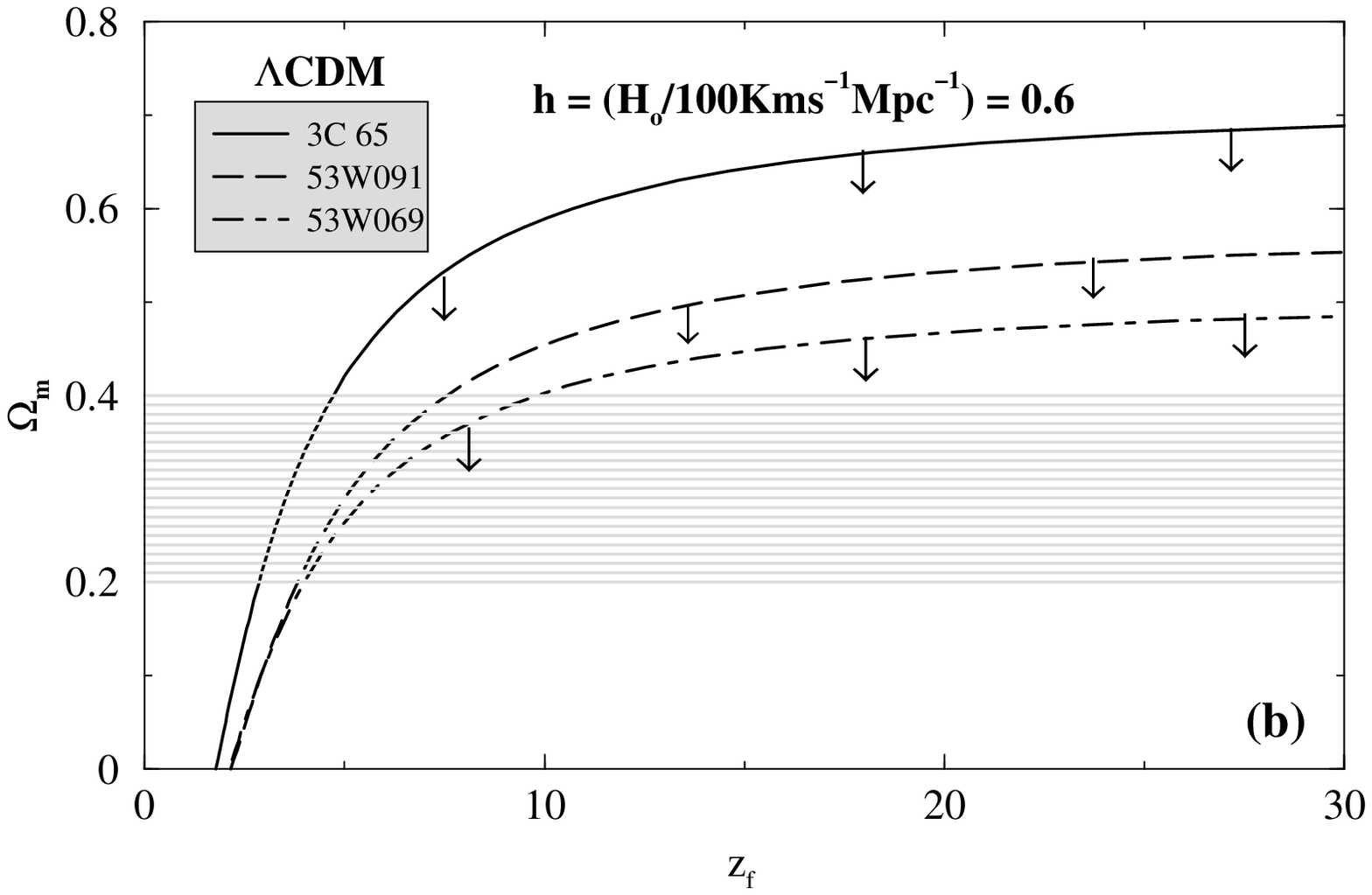,width=3.0truein,height=3.0truein}   
\hskip 0.1in} 
\caption{The $\Omega_m - z_f$ plane allowed by the  existence of OHRG's in 
the framework of OCDM (Panel a) and $\Lambda$CDM (Panel b). The shadowed
horizontal region corresponds to the observed range of $\Omega_{\rm{m}}$. 
The arrows delimit the available parameter space. The curves are also defined 
by the underestimated values of $t_g$ and the indicated lower limit of $H_o$.
For a given value of $\Omega_m$, we see that the most restrictive limit is
provided by the radio galaxy 53W069.}  
\end{figure}

\vspace{0.5in} 

\begin{figure} 
\vspace{.2in}  
\centerline{\psfig{figure=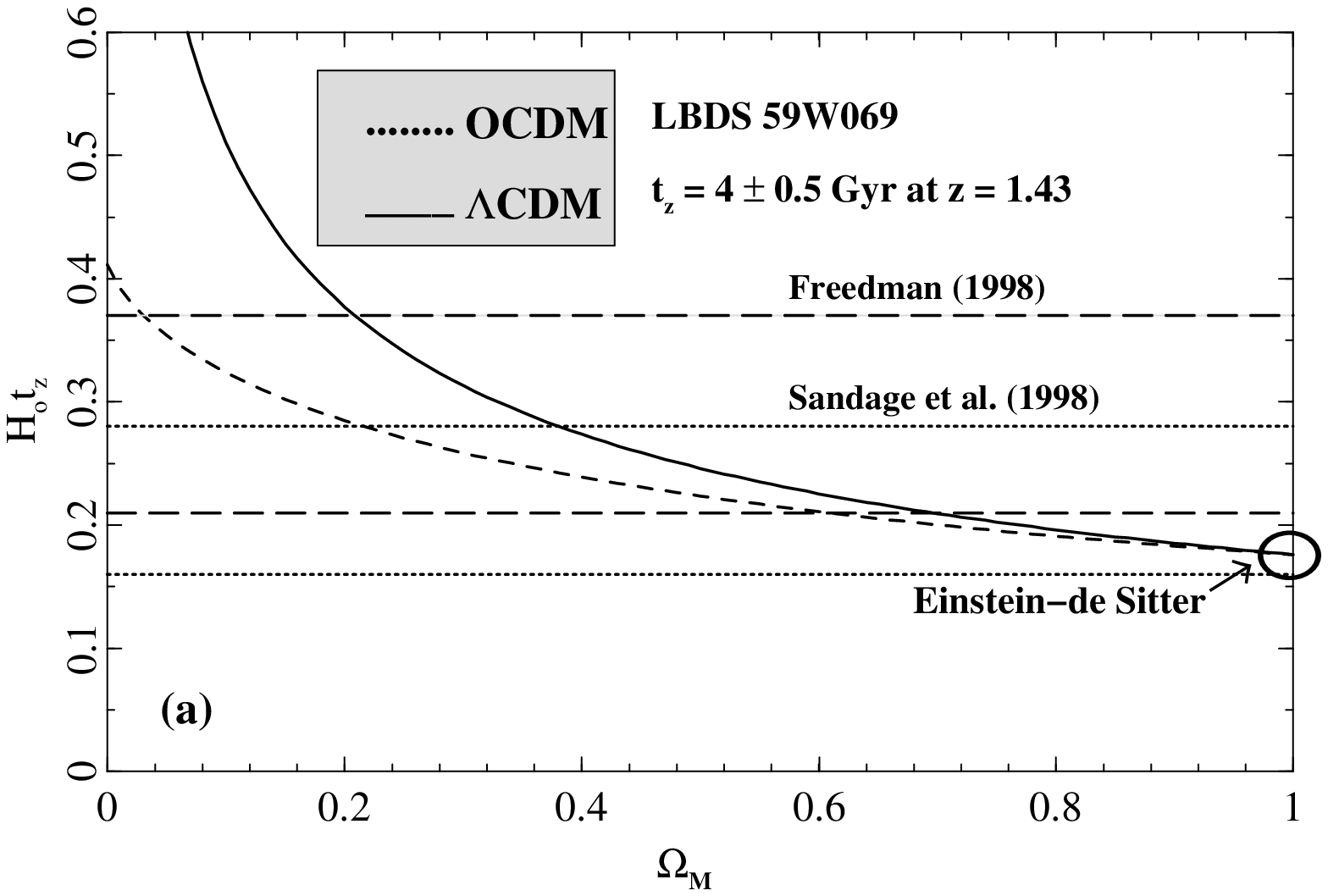,width=3.0truein,height=3.0truein}  
\psfig{figure=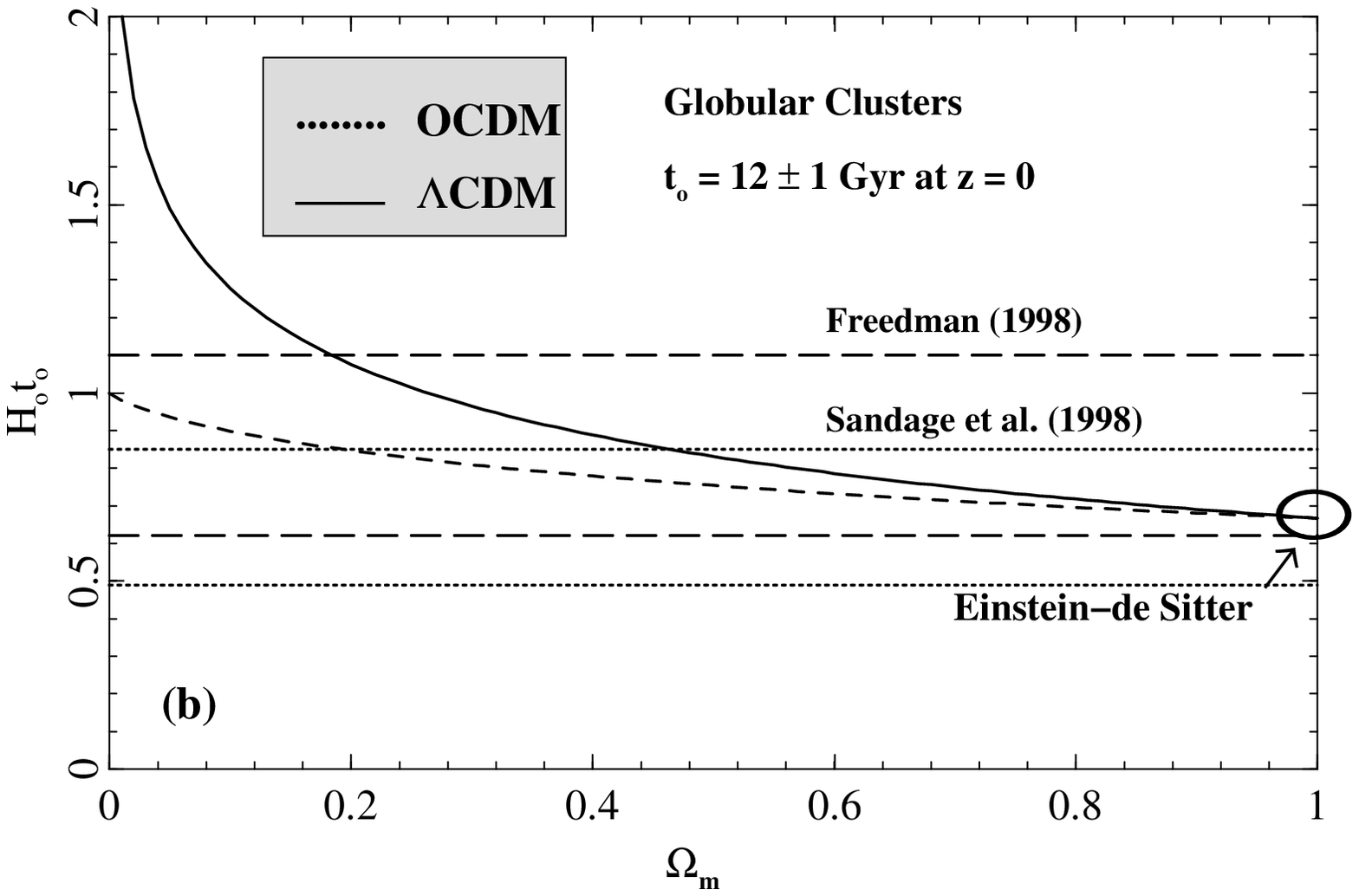,width=3.0truein,height=3.0truein}   
\hskip 0.1in}
\caption{$H_o t_z$ as a function of $\Omega_{\rm{m}}$. Dotted lines indicate the
$\pm 2\sigma$ limits of the age-parameter for $h = 0.55 \pm 0.05$  [27] whereas
dashed lines indicate the $\pm 2\sigma$ limits for $h = 0.7 \pm 0.1$, the
value obtained by the {\it HST Key Project} [18] and by other independent
estimates [17, 20]. Note that for $h = 0.7 \pm 0.1$ the Einstein-de Sitter
case (indicated in the panels) is off by 3$\sigma$ for the analysis of OHRGs.
For sake of completeness, Panel b shows similar analysis for globular clusters
- see also [19].} \end{figure}

\end{document}